\documentclass[abstract]{acmsiggraph}

\usepackage{mathptmx}
\usepackage{ifxetex}
\ifxetex
  \usepackage{fontspec}
\else
  \usepackage[utf8]{inputenc}
\fi

\usepackage{enumitem}

\usepackage{siunitx}
\ExplSyntaxOn
\NewDocumentCommand \hms { o > { \SplitArgument { 2 } { ; } } m }
  {
    \group_begin:
      \IfNoValueF {#1}
        { \keys_set:nn { siunitx } {#1} }
      \siunitx_hms_output:nnn #2
    \group_end:
  }
\cs_new_protected:Npn \siunitx_hms_output:nnn #1#2#3
  {
    \IfNoValueF {#1}
      {
        \tl_if_blank:nF {#1}
          {
            \SI {#1} { \hour }
            \IfNoValueF {#2} { ~ }
          }
      }
    \IfNoValueF {#2}
      {
        \tl_if_blank:nF {#2}
          {
            \SI {#2} { \minute }
            \IfNoValueF {#3} { ~ }
          }
      }
    \IfNoValueF {#3}
      { \tl_if_blank:nF {#3} { \SI {#3} { \second } } }
  }
\ExplSyntaxOff

\usepackage[acronym, shortcuts, nowarn]{glossaries}
\glsdisablehyper
\newacronym{av}{AV}{audiovisual}
\newacronym{ema}{EMA}{electromagnetic articulography}
\newacronym{ik}{IK}{inverse kinematics}
\newacronym{mri}{MRI}{magnetic resonance imaging}
\newacronym{tts}{TTS}{text-to-speech}
\makeglossaries

\title{Using multimodal speech production data to evaluate\\
  articulatory animation for audiovisual speech synthesis}
\author{\begin{tabular}{ccc}
Ingmar Steiner\thanks{\nolinkurl{ingmar.steiner@ucd.ie}} &
Korin Richmond\thanks{\nolinkurl{korin@cstr.ed.ac.uk}} &
Slim Ouni\thanks{\nolinkurl{slim.ouni@loria.fr}} \\
University College Dublin &
University of Edinburgh &
Université de Lorraine \\
\& Trinity College Dublin & & LORIA, UMR 7503 \\
\end{tabular}}
\pdfauthor{Ingmar Steiner, Korin Richmond, Slim Ouni}

\newcommand{\mngu}{{\tt mngu0}}

\begin{document}

\maketitle

\copyrightspace

\section{Introduction}

The importance of modeling speech articulation for high-quality \ac{av} speech synthesis is widely acknowledged.
Nevertheless, while state-of-the-art, data-driven approaches to facial animation can make use of sophisticated motion capture techniques, the animation of the intraoral articulators (viz.\ the tongue, jaw, and velum) typically makes use of simple rules or viseme morphing, in stark contrast to the otherwise high quality of facial modeling.
Using appropriate speech production data could significantly improve the quality of articulatory animation for \ac{av} synthesis.

\section{Articulatory animation}

To complement a purely data-driven \ac{av} synthesizer employing bimodal unit-selection \cite{Musti2011AVSP}, we have implemented a framework for articulatory animation \cite{Steiner2012IAST} using motion capture of the hidden articulators obtained through \ac{ema} \cite{Hoole2010SMC}.
One component of this framework compiles an animated 3D model of the tongue and teeth as an asset usable by downstream components or an external 3D graphics engine.
This is achieved by rigging static meshes with a pseudo-skeletal armature, which is in turn driven by the \ac{ema} data through \ac{ik}.
Subjectively, we find the resulting animation to be both plausible and convincing.
However, this has not yet been formally evaluated, and so the motivation for the present paper is to conduct an objective analysis.

\section{Multimodal speech production data}

\begin{figure}
  \centering
  \includegraphics[width=.4\linewidth]{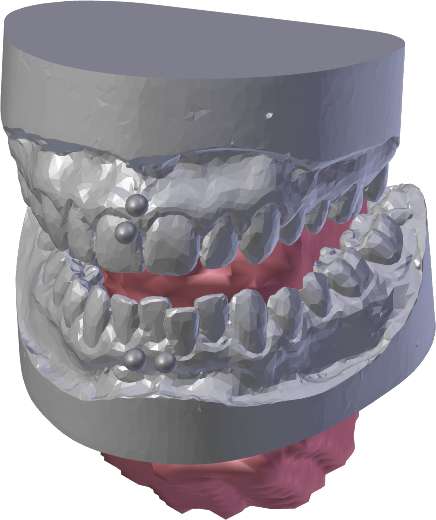}
  \qquad
  \includegraphics[width=.4\linewidth]{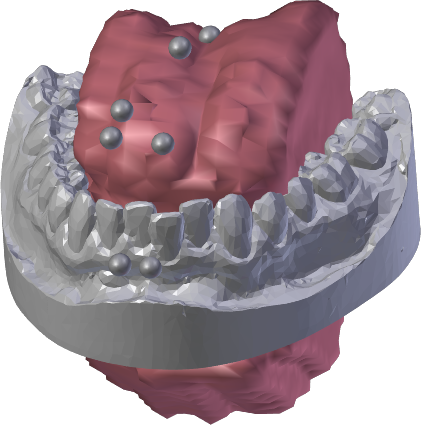}
  \caption{Animated articulatory model in bind pose, with and without maxilla; \ac{ema} coils rendered as spheres.}
  \label{fig:render}
\end{figure}

The \mngu\ articulatory corpus%
\footnote{freely available for research purposes from \url{http://mngu0.org/}}
contains a large set of 3D \ac{ema} data \cite{Richmond2011IS} from a male speaker of British English, as well as volumetric \ac{mri} scans of that speaker's vocal tract during sustained speech production \cite{Steiner2012JASA}.
Using the articulatory animation framework, static meshes of dental cast scans and the tongue (extracted from the \ac{mri} subset of the \mngu\ corpus) can be animated using motion capture data from the \ac{ema} subset, providing a means to evaluate the synthesized animation on the generated model (\autoref{fig:render}).

\section{Evaluation}

In order to analyze the degree to which the animated articulators match the shape and movements captured by the natural speech production data, several approaches are described.

\begin{itemize}[nosep]
  \item The positions and orientations of the \ac{ik} targets are dumped to data files in a format compatible with that of the 3D articulograph.
  This allows visualization and direct comparison of the animation with the original \ac{ema} data, using external analysis software.% (e.g., \cite{Ouni2012IS}).
  \item The distances of the \ac{ema}-controlled \ac{ik} targets to the surfaces of the animated articulators should ideally remain close to zero during deformation.
  Likewise, there should be collision with a reconstructed palate surface, but no penetration.
  \item A tongue mesh extracted from a volumetric \ac{mri} scan in the \mngu\ data, when deformed to a pose corresponding to a given phoneme, should assume a shape closely resembling the vocal tract configuration in the corresponding volumetric scan.
%  However, the difference in speaker posture across modalities may have a significant effect \cite{Kitamura2005AST}.
\end{itemize}

These evaluation approaches are implemented as unit and integration tests in the corresponding phases of the model compiler's build lifecycle, automatically producing appropriate reports by which the naturalness of the articulatory animation may be assessed.

% bibliography
\bibliographystyle{acmsiggraph}
\bibliography{ref}

\begin{thebibliography}{\protect\citename{Richmond et~al\mbox{.} }2011}

\bibitem[\protect\citename{Hoole and Zierdt }2010]{Hoole2010SMC}
{\sc Hoole, P., and Zierdt, A.}
\newblock 2010.
\newblock Five-dimensional articulography.
\newblock In {\em Speech Motor Control: New developments in basic and applied
  research}, B.~Maassen and P.~van Lieshout, Eds. Oxford University Press,
  331--349.

\bibitem[\protect\citename{Musti et~al\mbox{.} }2011]{Musti2011AVSP}
{\sc Musti, U., Colotte, V., Toutios, A., and Ouni, S.}
\newblock 2011.
\newblock Introducing visual target cost within an acoustic-visual
  unit-selection speech synthesizer.
\newblock In {\em Proc. 10th International Conference on Auditory-Visual Speech
  Processing (AVSP)}, 49--55.

\bibitem[\protect\citename{Richmond et~al\mbox{.} }2011]{Richmond2011IS}
{\sc Richmond, K., Hoole, P., and King, S.}
\newblock 2011.
\newblock Announcing the electromagnetic articulography (day 1) subset of the
  mngu0 articulatory corpus.
\newblock In {\em Proc. Interspeech}, 1505--1508.

\bibitem[\protect\citename{Steiner and Ouni }2012]{Steiner2012IAST}
{\sc Steiner, I., and Ouni, S.}
\newblock 2012.
\newblock Artimate: an articulatory animation framework for audiovisual speech
  synthesis.
\newblock In {\em Proc. {ISCA} Workshop on Innovation and Applications in
  Speech Technology}.

\bibitem[\protect\citename{Steiner et~al\mbox{.} }2012]{Steiner2012JASA}
{\sc Steiner, I., Richmond, K., Marshall, I., and Gray, C.~D.}
\newblock 2012.
\newblock The magnetic resonance imaging subset of the mngu0 articulatory
  corpus.
\newblock {\em Journal of the Acoustical Society of America 131}, 2 (Feb.),
  106--111.

\end{thebibliography}

\end{document}